\documentclass{aastex62}

\received{...}
\revised{...}
\accepted{...}

\submitjournal{ApJL}

\shorttitle{Failed multi-thermal flux rope eruption}
\shortauthors{Li et al.}

\begin{document}

%% title
\title{Failed Solar Eruption of A Multi-thermal Flux Rope}

%% authors
\correspondingauthor{Leping Li}
\email{lepingli@nao.cas.cn}

\author[0000-0001-5776-056X]{Leping Li}
\affil{National Astronomical Observatories, Chinese Academy of Sciences, Beijing 100101, Peoples's Republic of China}
\affiliation{University of Chinese Academy of Sciences, Beijing 100049, Peoples's Republic of China}

\author[0000-0001-5705-661X]{Hongqiang Song}
\affiliation{Shandong Provincial Key Laboratory of Optical Astronomy and Solar-Terrestrial Environment, and Institute of Space Sciences, Shandong University, Weihai, Shandong 264209, Peoples's Republic of China}

\author[0000-0001-9921-0937]{Hardi Peter}
\author[0000-0002-9270-6785]{Lakshmi Pradeep Chitta}
\affiliation{Max Planck Institute for Solar System Research, D-37077 G\"{o}ttingen, Germany}

%% abstract
\begin{abstract}

A magnetic flux rope (FR), hosting hot plasma, is thought to be central to the physics of coronal mass ejection.
Such FRs are widely observed with passbands of the Atmospheric Imaging Assembly (AIA) onboard the Solar Dynamics Observatory (SDO), that are sensitive to emission from the hot plasma around 10\,MK.
In contrast, observations of warmer (around 1\,MK) counterparts of FRs are sparse. 
In this study, we report the failed eruption of a multi-thermal FR, hosting both hot and warm plasma.
On 2015 May 1, a hot channel appeared in the AIA high temperature passbands out of the southeastern solar limb to the south of a nearby flare, and then erupted outward.
During the eruption, it rotated perpendicular to the erupting direction.
The hot channel stopped erupting, and disappeared gradually, showing a failed eruption.
During the hot channel eruption, a warm channel appeared sequentially in the AIA low temperature passbands.
It underwent the similar evolution, including the failed eruption, rotation, and disappearance, to the hot channel.
A bright compression front is formed in front of the warm channel eruption in AIA low temperature images.
Under the hot and warm channel eruptions, a small flare occurred, upon which several current sheets, connecting the erupting channels and the underneath flare, formed in the AIA high temperature passbands.
Investigating the spatial and temporal relation between the hot and warm channels, we suggest that both channels twist together, constituting the same multi-thermal FR that has plasma with the high and low temperatures.

\end{abstract}

%% keywords
%\keywords{Solar filaments (1495); Solar ultraviolet emission (1533); Plasma physics (2089); Solar corona (1483); Solar flares (1496); Solar magnetic reconnection (1504)}

%% introduction
\section{Introduction} \label{sec:int}

Solar magnetic flux rope (FR) is a coherent helical structure, often observed to form over polarity inversion lines (PILs) in active regions (ARs), with field lines wound around a central axis \citep{2010ApJ...718.1388D, 2016ApJ...818..148L, 2018SSRv..214...46G, 2020SSRv..216..131P}.
It often erupts explosively outward \citep{2001ApJ...562.1045K, 2014ApJ...789...46K, li16a}.
A successful FR eruption produces a coronal mass ejection \citep[CME;][]{2006ChJAA...6..247W,  2011ApJ...739...43L, li16b, 2013AdSpR..51.1967S}, a major driver of the space weather \citep{2010ApJ...722.1762L, 2014GeoRL..41.2673G, 2021LRSP...18....4T}. 
In a typical three-part CME structure, with a bright front, dark cavity, and bright core, the erupting FR is thought to be the centerpiece, i.e., the bright core \citep{2012A&A...539A...7L, 2019ApJ...883...43S, 2022ApJ...933...68S} or the dark cavity with the embedded bright core \citep[the erupting filament;][]{2011LRSP....8....1C}.
The failed FR eruption, however, has no CME associated \citep{2003ApJ...595L.135J, 2017ApJ...838...15L}.

Filament (or prominence) is composed of cooler, denser chromospheric material suspended in the hotter, rarer corona along PILs \citep{2006AdSpR..38..466Z, 2010SSRv..151..333M, 2013SoPh..282..147L, 2018ApJ...863..192L}.
It is commonly thought to be supported in magnetic dips of the FR \citep{2010ApJ...714..343G, 2014ApJ...780..130X, 2014ApJ...786L..16J, 2018ApJ...856..179Z}.
This is confirmed observationally by the filament rotation around its central axis during the eruption \citep{2014ApJ...787...11J, 2014ApJ...797...52Y, 2016NatCo...711837X}.
Moreover, the erupting filament is heated to higher temperatures, and seen in transition region and coronal diagnostics, e.g., in extreme ultraviolet (EUV), showing the multi-thermal FR \citep{2003ApJ...598..683D, 2013AJ....145..153Y, 2022ApJ...932L...9K}.

Employing Atmospheric Imaging Assembly \citep[AIA;][]{2012SoPh..275...17L} images on board the Solar Dynamics Observatory \citep[SDO;][]{2012SoPh..275....3P}, \citet{2011ApJ...732L..25C} and \citet{2012NatCo...3..747Z} found the coherently elongated hot channels (or hot blobs), and identified them as a type of new and promising evidence of FRs.
The off-limb hot channel shows a writhed channel-like structure with two elbows inclining to the opposite directions and the middle being concaved toward the surface in AIA 131 and 94\,\AA~images, and appears as a dark cavity in other low temperature, e.g., 171\,\AA, passbands \citep[see a review in][and references therein]{2017ScChD..60.1383C}. 
\citet{2012ApJ...761...62C} quantified the temperature and emission measure (EM) of hot channels using the differential EM (DEM) analysis, and found that the emission of hot channels is from a broad temperature range, with a DEM-weighted average temperature larger than $\sim$8\,MK.
Investigating the temperature evolution of an erupting hot channel, \citet{2014ApJ...784...48S} reported that the average temperature of hot channel increases from 5 to 10\,MK steadily during the early phase. 
Recently, hundreds of hot channels have been detected in AIA high temperature images \citep{2015ApJ...808..117N, 2020A&A...642A.109N, 2015A&A...580A...2Z}.

Analyzing the failed and successful eruptions, \citet{2014ApJ...789L..35C} and \citet{2014ApJ...794..149C} investigated the relationship between the hot channel and the associated filament, respectively.
They suggested that the hot channel hosts the filament near its bottom.
Furthermore, after the appearance of hot channel in AIA high temperature images, \citet{2014ApJ...794..149C} reported that the FR is seen in AIA low temperature images when the low-lying filament appears, indicating the multi-thermal nature of the FR.
As the FR quickly erupted, evolution of the multi-thermal structures is difficult to analyze.
Employing observations from three viewing angles, i.e., the SDO and Solar Terrestrial Relations Observatory (STEREO) A and B, \citet{2013A&A...552L..11L} studied an eruption of two FRs with no associated filament, and presented that the FRs have both hot and cool components.
However, both components cannot be detected with one instrument alone.
The hot channels appear only in SDO/AIA high temperature images, while the warm channels are observed solely in STEREO/EUV Imager (EUVI) low temperature images.

In this Letter, investigating the failed eruption of a FR with no associated filament, we report the multi-thermal nature of the FR, including both hot and warm plasma.
The observations, results, and summary and discussion are shown in Sections\,\ref{sec:obs}-\ref{sec:sum}, respectively.

%% observations
\section{Observations}\label{sec:obs}

SDO/AIA is a set of normal-incidence imaging telescopes, acquiring solar atmospheric images in 10 wavelength passbands.
Different AIA passbands show plasma at different temperatures, e.g., 131\,\AA~peaks at $\sim$10 MK (Fe XXI) and $\sim$0.6\,MK (Fe VIII), 94\,\AA~peaks at $\sim$7.2\,MK (Fe XVII), 335\,\AA~peaks at $\sim$2.5\,MK (Fe XVI), 211\,\AA~peaks at $\sim$1.9\,MK (Fe XIV), 193\,\AA~peaks at $\sim$1.5\,MK (Fe XII), 171\,\AA~peaks at $\sim$0.9\,MK (Fe IX), and 304\,\AA~peaks at $\sim$0.05\,MK (He II).
In this study, we employ the AIA 131, 94, 335, 211, 193, 171, and 304\,\AA~images, with spatial sampling and time cadence of 0.6\arcsec\,pixel$^{-1}$ and 12\,s, to investigate the failed multi-thermal FR eruption.
To better show the evolution, the AIA images are enhanced by employing the Multiscale Gaussion Normalization (MGN) technique \citep{2014SoPh..289.2945M}.

%% Results
\section{Results}\label{sec:res}

\subsection{Failed eruption of a hot channel}\label{sec:hce}

On 2015 May 1, the NOAA AR 12335, located at heliographic position E76\,S14, was observed by the SDO.
In this AR, a C2.3 flare occurred from 02:48 to 03:07 UT, peaking at  02:57 UT; see Figure\,\ref{f:general}(a) and the red curves in Figure\,\ref{f:timeslices}.
During the flare, to the south of the AR, a hot channel, representing the FR, appeared out of the southeastern limb in AIA 94 and 131\,\AA~images; see Figures\,\ref{f:general}(a) and (b).
It then erupted toward the southeast.
During the eruption, a twisted structure of the hot channel is identified; see Figure\,\ref{f:general}(b).

Using six AIA passbands, including 94, 335, 211, 193, 171, and 131\,\AA, we analyze the temperature and EM of the hot channel.
Here, we employ the DEM analysis using ``xrt\_dem\_iterative2.pro", identical to \citet{2012ApJ...761...62C}. 
In the early phase of the eruption, e.g., at 04:21 UT, the hot channel region, enclosed by the red rectangle in Figure\,\ref{f:general}(b), is chosen to calculate the DEM.
The region out of the hot channel, enclosed by the purple rectangle in Figure\,\ref{f:general}(b), is chosen for the background emission that is subtracted from the hot channel region.
In each region, the digital number counts in each of six AIA passbands are temporally normalized by the exposure time and spatially averaged over all pixels.
The DEM curve of the hot channel region is displayed in Figure\,\ref{f:dems}(a).
The DEM-weighted average temperature and EM are 8.4 MK and 6.2$\times$10$^{26}$\,cm$^{-5}$, respectively.
Employing the EM, the density ($n_{e}$) of the hot channel is estimated using $n_{e}$=$\sqrt{\frac{EM}{D}}$, where $D$ is the line-of-sight (LOS) depth of the hot channel.
Assuming that the depth ($D$) equals the width ($W$) of the hot channel, then the density is $n_{e}$=$\sqrt{\frac{EM}{W}}$.
For the hot channel width, first we get the intensity profile in the 94\,\AA~passband perpendicular to the hot channel. 
We use the mean intensity surrounding the hot channel as the background emission, and subtract it from the intensity profile.
Employing a single Gaussian, we fit the residual intensity profile, and achieve the FWHM of the single Gaussian fit as the width.
The hot channel width, where we calculate the DEM, is $\sim$13\,Mm.
Using EM=6.2$\times$10$^{26}$\,cm$^{-5}$ and $W$=13\,Mm, we obtain the hot channel density to be 6.9$\times$10$^{8}$\,cm$^{-3}$.

Along the erupting direction; see the blue line AB in Figure\,\ref{f:general}(e), a time slice of AIA 94\,\AA~images is made, and displayed in Figure\,\ref{f:timeslices}(a1).
The left pink arrow in Figure\,\ref{f:timeslices}(a1) marks the appearance of hot channel during the nearby flare, while the right one denotes the eruption, with an acceleration and speed of $\sim$3.5\,m\,s$^{-2}$ and $\sim$55\,km\,s$^{-1}$.  
Moreover, the rotation of hot channel perpendicular to the erupting direction is identified; see the online animated version of Figure\,\ref{f:7wls}.
Along the rotating direction at two heights; see the purple and green lines CD and EF in Figures\,\ref{f:general}(e) and (f), two time slices of AIA 94\,\AA~images are made, and shown in Figures\,\ref{f:timeslices}(a2) and (a3), respectively.
The pink arrows in Figure\,\ref{f:timeslices}(a2) mark the rotation, with a speed of $\sim$20\,km\,s$^{-1}$.
Figure\,\ref{f:timeslices}(a3) displays the further expansion of hot channel due to the rotation.
The hot channel then evolved into an $\alpha$ shape, with the top middle being concaved toward the surface; see Figures\,\ref{f:general}(g) and \ref{f:7wls}(b).
In the late phase of the eruption, e.g., at 05:36 UT, the temperature and EM of the hot channel region are also calculated; see the red and purple rectangles in Figure\,\ref{f:general}(g).
The DEM curve is displayed in Figure\,\ref{f:dems}(b).
Compared with that in Figure\,\ref{f:dems}(a), more emission of the hot channel from low temperature plasma is detected.
The DEM-weighted average temperature and EM are separately 6.8\,MK and 5.2$\times$10$^{26}$\,cm$^{-5}$.
Measuring the width ($\sim$21\,Mm), we estimate the density of hot channel to be 5.0$\times$10$^{8}$\,cm$^{-3}$, less than that in the early phase.
This may be caused by the expansion of the erupting hot channel.
The hot channel stops erupting, and disappears gradually, suggesting a failed eruption.

\subsection{Failed eruption of a warm channel}\label{sec:wce}

After the hot channel eruption began, an erupting structure appeared sequentially in AIA low temperature, e.g., 335, 211, 193, 171, and 304\,\AA, images; see Figures\,\ref{f:general}(c)-(e).
It, termed warm channel, has two legs and finer structure, denoted by the green solid arrows in Figures\,\ref{f:general}(c) and (d), and \ref{f:general}(e) and \ref{f:7wls}(e)-(f), respectively.
During the eruption, a weaker signature of the warm channel was observed in 304\,\AA~images for a short time ($\sim$1\,hr), but not detected in H$\alpha$ images.
No filament is thus associated with the eruption.
Similar to the hot channel, the temperature and EM of the warm channel region are calculated in the early phase of the eruption; see the red and pink rectangles in Figures\,\ref{f:7wls}(a)-(f).
The DEM curve is displayed in Figure\,\ref{f:dems}(c).
It shows that the emission is from both hot and warm plasma, as the chosen region also includes the hot channel; see Figure\,\ref{f:7wls}(b).
The DEM-weighted average temperature and EM are 1.9\,MK and 1.2$\times$10$^{27}$\,cm$^{-5}$, respectively.
The warm channel thus has more emission from low temperature plasma. 
We measure the width ($\sim$9.5\,Mm) of the warm channel, and estimate the density to be 1.1$\times$10$^{9}$\,cm$^{-3}$, larger than that of the hot channel.

In front of the eruption, a bright compression front forms in AIA low temperature, e.g., 211 and 193\,\AA, images; see Figures\,\ref{f:general}(c) and (d).
The temperature and EM of the compression front region are calculated; see the green and blue rectangles in Figure\,\ref{f:general}(c).
The DEM curve is shown in Figure\,\ref{f:dems}(e), with the DEM-weighted average temperature and EM of 2.0\,MK and 2.8$\times$10$^{26}$\,cm$^{-5}$, respectively. 
The emission of compression front mainly originates from low temperature plasma.
We measure the width ($\sim$14\,Mm) of the compression front, and estimate the density to be 4.5$\times$10$^{8}$\,cm$^{-3}$, less than those of the hot and warm channels.
Moreover, the nearby loops are pushed outward by the warm channel eruption, and then retracted; see Figures\,\ref{f:general}(c)-(e) and the online animated version of Figure\,\ref{f:7wls}.

Along the erupting direction; see the blue line AB in Figure\,\ref{f:general}(e), time slices of AIA 211 and 171\,\AA~images are made, and displayed in Figures\,\ref{f:timeslices}(b1) and (c1), respectively.
In these figures, the blue arrows denote the formation and propagation of the compression front, with an acceleration and speed of $\sim$1.9\,m\,s$^{-2}$ and $\sim$22\,km\,s$^{-1}$.
The green arrows mark the warm channel eruption, with a speed of $\sim$27\,km\,s$^{-1}$.
The warm channel stops erupting slowly with a deceleration of $\sim$2.3\,m\,s$^{-2}$, and disappears gradually; see Figure\,\ref{f:timeslices}(c1).
The warm channel eruption is thus a failed one, consistent with the hot channel eruption.
The temperature and EM of the warm channel region in the late phase of the eruption is calculated; see the red and pink rectangles in Figure\,\ref{f:general}(f).
The DEM curve is displayed in Figure\,\ref{f:dems}(d).
Compared with that in Figure\,\ref{f:dems}(c), the emission of warm channel is mainly from low temperature plasma.
The DEM-weighted average temperature and EM are 2.0\,MK and 3.2$\times$10$^{26}$\,cm$^{-5}$, respectively.
Although the DEM curves in Figures\,\ref{f:dems}(c) and (d) are different, they have the similar DEM-weighted average temperatures, i.e., 1.9 and 2.0 MK. This is because the emission of warm channel in the early and late phases of the eruption is both dominated by plasma with low temperatures (5.8$\leqslant$log$T$$\leqslant$6.7).
We measure the width (3.6\,Mm) of the warm channel, and estimate the density to be 5.4$\times$10$^{8}$\,cm$^{-3}$, about half of that in the early phase.
This may be due to the expansion of the erupting warm channel.

Consistent with the hot channel, during the eruption, the warm channel also rotated perpendicular to the erupting direction, with a rotating angle of $\sim$90$^{\circ}$.
Along the rotating direction; see the purple (green) line CD (EF) in Figure\,\ref{f:general}(e (f)), time slices of AIA 211 and 171\,\AA~images are made, and shown in Figures\,\ref{f:timeslices}(b2 (b3)) and (c2 (c3)), respectively.
In these figures, the red and black arrows separately mark the expansion and retraction of the surrounding loops, with mean speeds of $\sim$1.5 and $\sim$8.5\,km\,s$^{-1}$.
The green arrows denote two warm channel legs, that move across each other with a speed of $\sim$1.5\,km\,s$^{-1}$, indicating the rotation.
The cyan arrows in Figures\,\ref{f:timeslices}(b1)-(c1) show one warm channel leg caused by the rotation.
In Figures\,\ref{f:timeslices}(b3) and (c3), the red and black arrows also denote the expansion and retraction of the surrounding loops.
The blue arrows denote the passage of the compression front, and the green arrows show the passage and rotation of the warm channel.

\subsection{Relation between the failed eruptions of hot and warm channels}\label{sec:relation}

Employing AIA multi-wavelength images, the spatial and temporal relation between the failed eruptions of hot and warm channels is investigated.
Figures\,\ref{f:general}(h) and \ref{f:7wls}(h) show the composite images of AIA 94, 211, and 171\,\AA~images.
The timeslices in Figures\,\ref{f:timeslices}(d1)-(d3) separately display the composites of timeslices in Figures\,\ref{f:timeslices}(a1)-(c1), (a2)-(c2), and (a3)-(c3).
Figure\,\ref{f:timeslices}(d1) shows that the compression front is formed in front of both eruptions, which have identical erupting trajectory.
In the late phase of the eruptions, the warm channel overtook the hot channel before they stopped erupting; see Figure\,\ref{f:general}(h). 
During the eruptions, both channels rotated similarly perpendicular to the erupting direction; see Figures\,\ref{f:timeslices}(d2) and (d3).
All the results indicate that both channels belong to the same FR, i.e., a multi-thermal FR with high and low temperature plasma.

Underneath the erupting FR, a small flare takes place; see Figure\,\ref{f:general}(g), and the post-flare loops are seen sequentially in AIA images.
Above the flare, several ray-like structures appear in AIA 94 and 131\,\AA~images; see the white rectangle in Figure\,\ref{f:general}(g). 
They represent current sheets, connecting the erupting FR and the flare \citep{li16b}.
Besides the retracting loops, other loops, marked by the blue arrow in Figure\,\ref{f:7wls}(d), moving toward the current sheets with a speed of $\sim$8.5 km\,s$^{-1}$, are identified, showing the reconnection inflows.
Using the Alfv\'{e}n speed (500-3000\,km\,s$^{-1}$) in the corona as the reconnection outflow speed, we estimate the reconnection rate to be 0.003-0.017.
The temperature and EM of the current sheet region is calculated; see the purple and black rectangles in Figure\,\ref{f:general}(g).
The DEM curve is shown in Figure\,\ref{f:dems}(f), with the DEM-weighted average temperature and EM of 12.3\,MK and 4.5$\times$10$^{27}$\,cm$^{-5}$.
We measure the width ($\sim$3.6\,Mm) of the current sheet, and estimate the density to be 3.5$\times$10$^{9}$\,cm$^{-3}$, larger than that of the warm channel.

%% Summary and discussion
\section{Summary and discussion}\label{sec:sum}

Employing SDO/AIA images, we investigate the failed eruption of multi-thermal FR.
On 2015 May 1, a C2.3 flare occurred in AR 12335 near the southeastern limb.
To the south, the hot channel appeared in AIA 94 and 131\,\AA~images during the flare, and then erupted.
During the eruption, it rotated perpendicular to the erupting direction.
The hot channel stopped erupting, showing the failed eruption, and then gradually disappeared.
During the hot channel eruption, the warm channel appeared sequentially in AIA 335, 211, 193, 171, and 304\,\AA~images.
Similar to the hot channel, it also erupted, rotated, stopped erupting, and disappeared.
In front of the warm channel eruption, the compression front was seen in AIA low temperature images.
The relation between the hot and warm channel eruptions suggests that both channels belong to the same FR, showing plasma with high and low temperatures.
Underneath the erupting FR, a small flare occurred, upon which several current sheets, connecting the FR and the flare, formed.

\begin{figure}[ht!]
\centering
\plotone{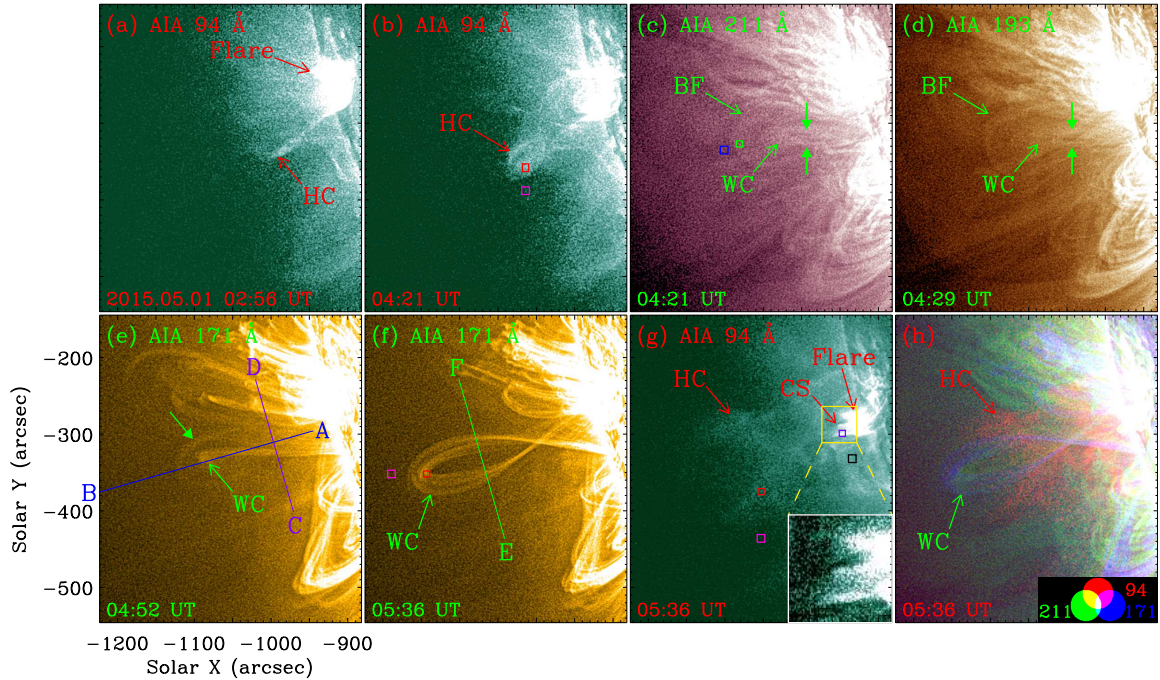}
\caption{General information of the failed multi-thermal FR eruption. 
(a)-(b) and (g) SDO/AIA 94, (c) 211, (d) 193, and (e)-(f) 171\,\AA~images, enhanced by the MGN technique, and (h) a composite of 94, 211, and 171\,\AA~images. 
The red rectangles in (b), (f), and (g), and green and purple rectangles in (c) and (g) enclose regions for the DEM curves in Figures\,\ref{f:dems}(a), (d), (b), (e) and (f). 
The pink rectangles in (b), (f), and (g), and blue and black rectangles in (c) and (g) mark the locations where the background emissions are measured.
The green solid arrows in (c)-(e) denote the warm channel fine structures.
The blue, purple, and green lines AB, CD, and EF in (e) and (f) separately show the positions of time slices in Figures\,\ref{f:timeslices}(a1)-(d1), (a2)-(d2), and (a3)-(d3). 
In (g), the AIA 94\,\AA~image in the yellow rectangle is enlarged in the white rectangle.
The HC, WC, BF, and CS mean the hot channel, warm channel, bright front, and current sheet, respectively.
\label{f:general}}
\end{figure}

\begin{figure}[ht!]
\plotone{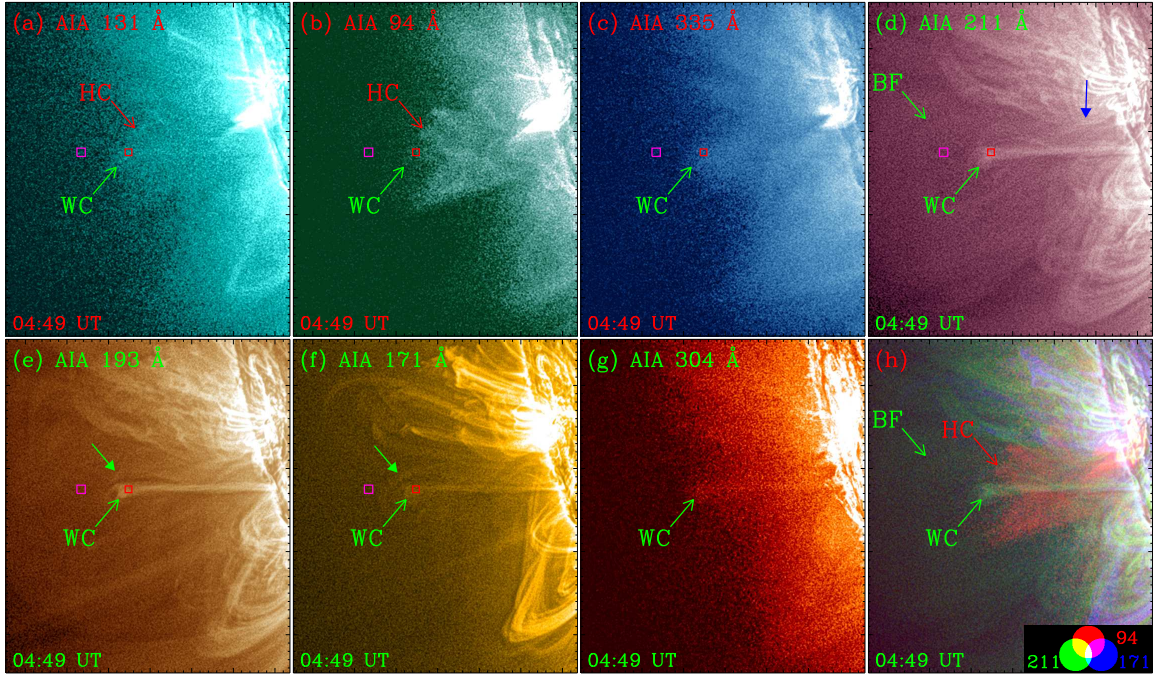}
\centering
\caption{Failed multi-thermal FR eruption observed by AIA. 
(a)-(g) AIA 131, 94, 335, 211, 193, 171, and 304\,\AA~images, enhanced by the MGN technique, and (h) a composite of 94, 211, and 171\,\AA~images, with the same FOV as Figure\,\ref{f:general}.
The red and pink rectangles in (a)-(f) separately enclose the region for the DEM curve in Figure\,\ref{f:dems}(c), and the location where the background emission is measured.
The blue and green solid arrows in (d)-(f) mark the loops and warm channel finer structure.
An animation of the unannotated AIA observations is available. 
It covers 8\,hr starting at 02:00\,UT, with a time cadence of 1\,minute.
(An animation of this figure is available.)
\label{f:7wls}}
\end{figure}

\begin{figure}[ht!]
\plotone{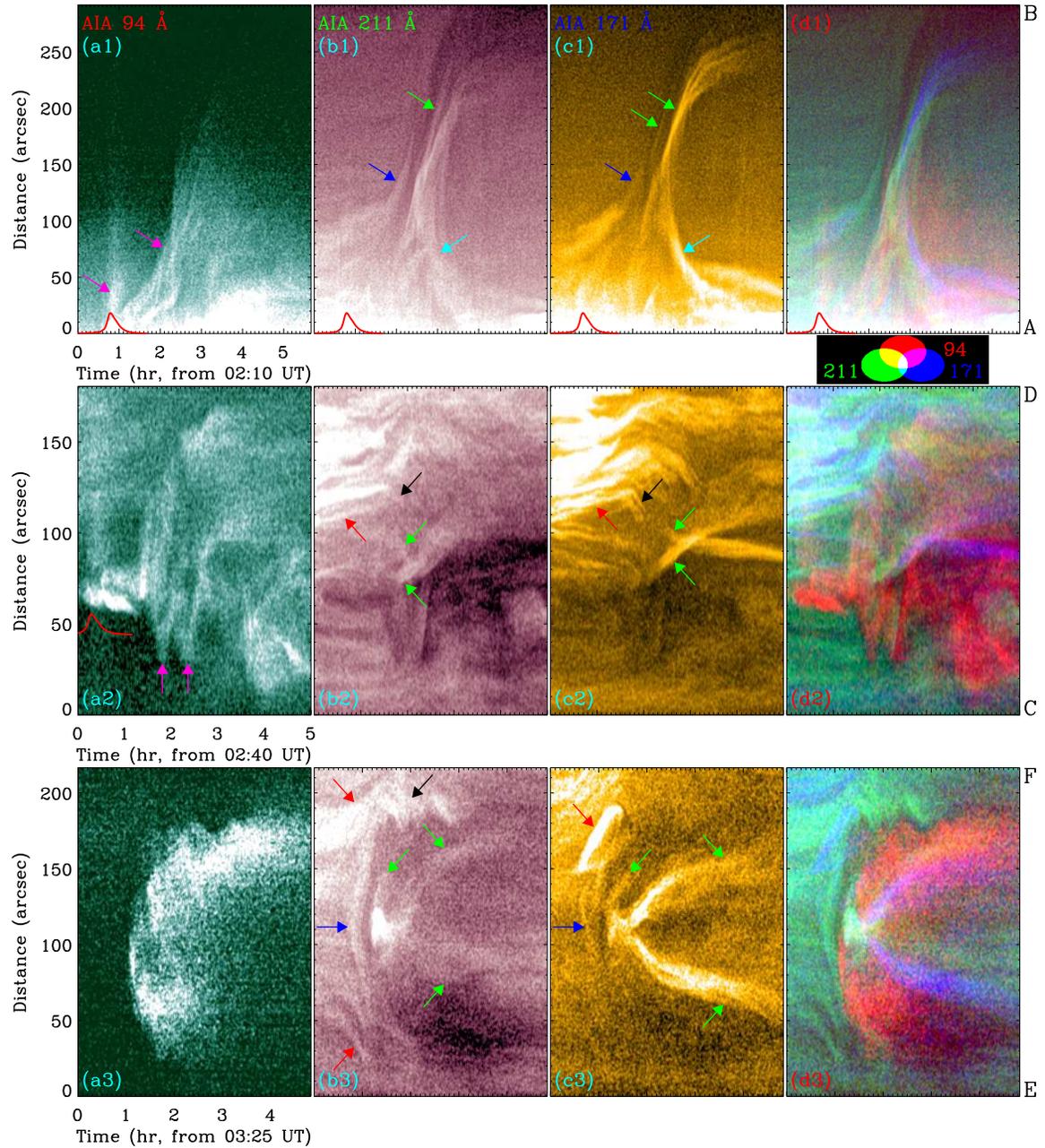}
\centering
\caption{Temporal evolution of the failed FR eruption.
(a1)-(d1), (a2)-(d2), and (a3)-(d3) Time slices of AIA 94, 211, and 171\,\AA~images and their composites along the blue, purple, and green lines AB, CD, and EF in Figures\,\ref{f:general}(e) and (f), respectively. 
The red lines in (a1)-(d1) and (a2) show the GOES-15 1-8\,\AA~soft X-ray flux.
The pink arrows in (a1)-(a2) mark the hot channel.
The blue arrows in (b1)-(c1) and (b3)-(c3) denote the compression front.
The green and cyan arrows in (b1)-(b3) and (c1)-(c3) mark the warm channel.
The red and black arrows in (b2)-(c2) and (b3)-(c3) denote the loops. 
\label{f:timeslices}}
\end{figure}

\begin{figure}[ht!]
\centering
\includegraphics[width=0.96\textwidth]{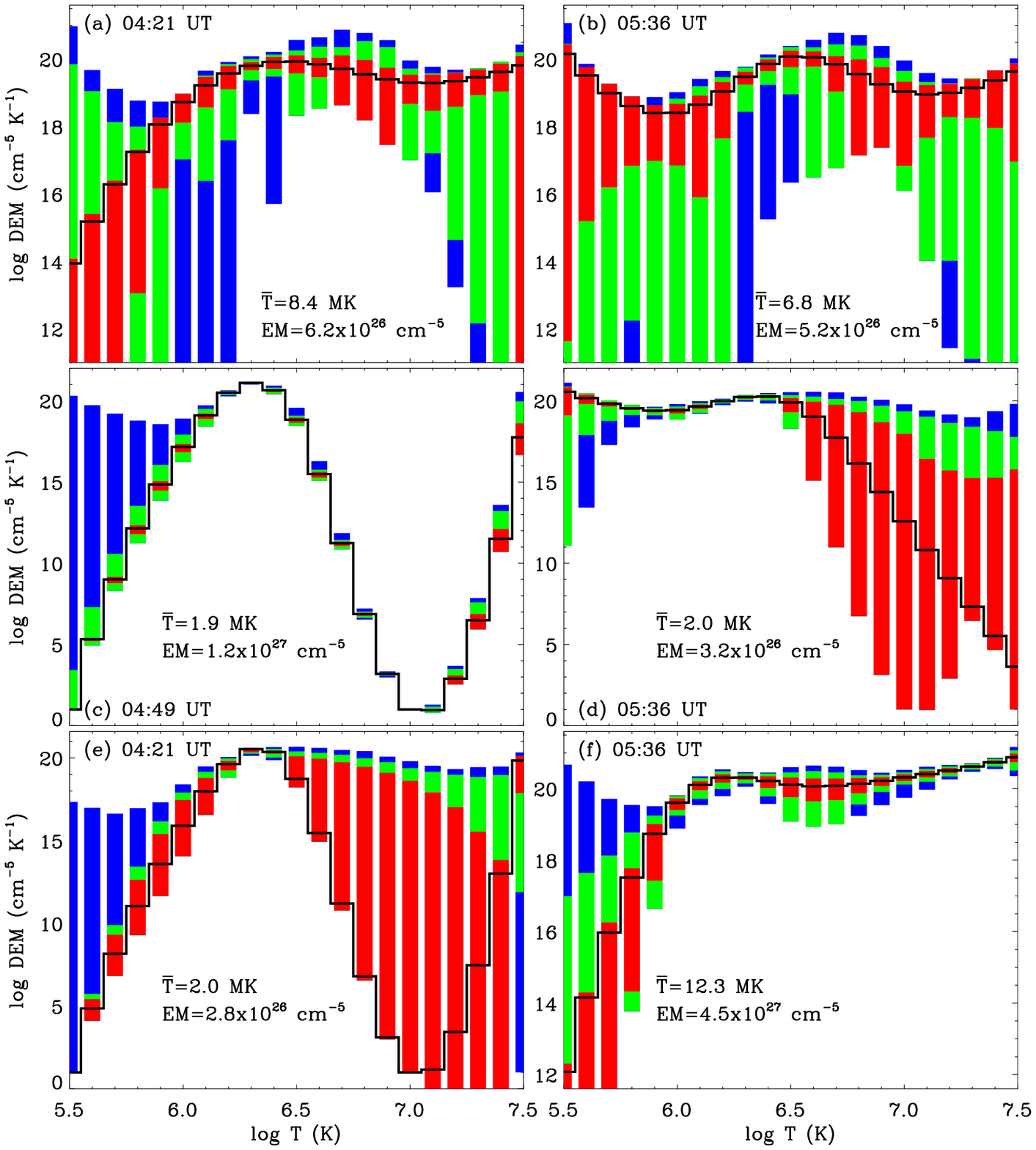}
\caption{Temperature and EM of the failed FR eruption.
(a)-(d) DEM curves for the FR regions enclosed by the red rectangles separately in Figures\,\ref{f:general}(b), \ref{f:general}(g), \ref{f:7wls}(a)-(f), and \ref{f:general}(f). 
(e)-(f) Similar to (a)-(d), but for the compression front and current sheet regions enclosed by the green and purple rectangles in Figures\,\ref{f:general}(c) and (g).
The black lines are the best-fit DEM distributions, and the red, green, and blue rectangles separately represent the regions containing 50\%, 51\%-80\%, and 81\%-95\% of the Monte Carlo solutions. 
\label{f:dems}}
\end{figure}

A multi-thermal FR with both hot and warm plasma is reported.
The hot channel appears in AIA high temperature images, which may be activated by the nearby flare, according to their spatial and temporal relationship.
The property of the hot channel, and also the compression front, is consistent with those previously investigated \citep{2012ApJ...761...62C, 2015ApJ...808..117N}.
Associated with the hot channel, the warm channel, with a lower temperature and higher density, is observed.
Different from that in \citet{2014ApJ...794..149C}, it has no associated filament, and appears in more low temperature passbands.
Although the general evolution, e.g., eruption and rotation, of both channels is similar, some different evolution between them is identified. 
The hot and warm channels hence twist together, with different topologies.
Moreover, the warm channel in Figure\,\ref{f:general}(f) looks like the cool compression front with the underneath dark cavity, that encloses the hot FR \citep{2012ApJ...761...62C, 2012NatCo...3..747Z}. However, this impression is not supported by the composite image in Figure\,\ref{f:general}(h) which shows that much of the hot structure surrounds the cooler structure instead of being embedded into it. This result provides an extra proof of the multi-thermal nature of the FR.

Failed eruption of the multi-thermal FR is presented.
The rotation of the erupting FR indicates that the eruption may be driven by the kink instability of FR \citep{2003ApJ...595L.135J}.
In deciding whether the eruption is failed or successful, the decrease of overlying field with height plays an important role  \citep{2005ApJ...630L..97T}.
The formation and propagation of the compression front show that overlying loops exist above the FR.
The small flare underneath the eruption hardly decreases the overlying field by reconnection between two legs of overlying loops \citep{2011RAA....11..594S}.
The overlying loops thus may confine the erupting FR \citep{2013ApJ...778...70C}.
Moreover, the FR rotation, with a larger rotating angle of $\sim$90$^{\circ}$, may also be responsible for the failed eruption \citep{2019ApJ...877L..28Z}.
The apparent rotation of the FR observed here is reminiscent of the early phases of erupting FRs that are kink-unstable \citep{2005ApJ...630L..97T}.
As the failed FR eruption is an off-limb event, it is difficult to determine and characterize the nature of kink and torus instabilities in this case. To understand the physical mechanism of the FR eruption, more observations are needed.

%% figures

%% acknowledgements
\acknowledgments

The authors thank the referee for helpful comments that led to improvements in the manuscript. We are indebted to the SDO team for providing the data. 
AIA images are the courtesy of NASA/SDO and the AIA, EVE, and HMI science teams. 
This work is supported by the Strategic Priority Research Program of Chinese Academy of Sciences (CAS), grant No. XDB 41000000, the National Natural Science Foundations of China (12073042 and U2031109), the Key Research Program of Frontier Sciences of CAS (ZDBS-LY-SLH013), and Yunnan Academician Workstation of Wang Jingxiu (No. 202005AF150025). 
L.P.C. gratefully acknowledges funding by the European Union. Views and opinions expressed are however those of the author(s) only and do not necessarily reflect those of the European Union or the European Research Council (grant agreement No. 101039844). Neither the European Union nor the granting authority can be held responsible for them.

%% references

\end{document}